\documentclass[]{aa} % for the letters 
\usepackage{graphicx}
\usepackage{color}

\usepackage{txfonts}
\def\spi{\mbox{INTEGRAL/SPI}}
\def\al{\mbox{$^{26}$Al}}
\def\gtsim {>\kern-1.2em\lower1.1ex\hbox{$\sim$}~}   % Greater than sim
\def\ltsim {<\kern-1.2em\lower1.1ex\hbox{$\sim$}~}   % Less than sim

\begin{document}

\title{The aluminium-26 distribution in a cosmological simulation \\ of a Milky Way-type Galaxy}
\titlerunning{\al ~in a chemodynamical simulation}

 \author{B. {Wehmeyer}
          \inst{1,2,3,4}\fnmsep\thanks{benjamin.wehmeyer@uwr.edu.pl}
          \and
          C. {Kobayashi}\inst{4}
          \and
          A. {Yag\"ue L\'opez}\inst{5}
          \and
          M. {Lugaro}\inst{2,3,6,7}
          }

\institute{
Institute of Theoretical Physics, University of Wroc{\l}aw, 50-204 Wroc{\l}aw, Poland
\and
Konkoly Observatory, HUN-REN Research Centre for Astronomy and Earth Sciences, Konkoly-Thege Mikl\'os út~15-17, Budapest~1121, Hungary
\and
CSFK, MTA Centre of Excellence, Konkoly-Thege Mikl\'os út~15-17, Budapest~1121, Hungary
\and
Centre for Astrophysics Research, University of Hertfordshire, College Lane, Hatfield AL10 9AB, UK
\and
Computer, Computational and Statistical Sciences (CCS) Division, Center for Theoretical Astrophysics, Los Alamos National Laboratory, Los Alamos, NM 87545, USA
\and
ELTE E\"{o}tv\"{o}s Lor\'and University, Institute of Physics and Astronomy, P\'azm\'any P\'eter s\'et\'any 1/A, Budapest~1117, Hungary
\and
School of Physics and Astronomy, Monash University, VIC 3800, Australia\\
}

\date{Received 18 Aug 2024 / accepted 23 Feb 2025}

% \abstract{}{}{}{}{} 
% 5 {} token are mandatory
 
  \abstract
   {The 1.8 MeV $\gamma$-rays corresponding to the decay of the radioactive isotope \al{} (with a half-life of $0.72$~Myr) have been observed by the SPI detector on the INTEGRAL spacecraft and extensively used as a tracer of star formation and current nucleosynthetic activity in the Milky Way Galaxy. Further information is encoded in the observation related to the higher \al{} content found in  regions of the Galaxy with the highest line-of-sight (LoS) velocity relative to an observer located in the Solar System. However, this feature remains unexplained.

   }
   {We ran a cosmological "zoom-in" chemodynamical simulation of a Milky Way-type galaxy, including the production and decays of radioactive nuclei in a fully self-consistent way.
   We then analyzed the results to follow the evolution of \al{} throughout the lifetime of the simulated galaxy to provide a new method for interpreting the \al{} observations.}
   {We included the massive star sources of \al{} in the Galaxy and its radioactive decay into a state-of-the-art galactic chemical evolution model, coupled with cosmological growth and hydrodynamics. This approach allowed us to follow the spatial and temporal evolution of the \al{} content in the simulated galaxy.}
   {Our results are in agreement with the observations with respect to the fact that gas particles in the simulation with relatively higher \al{} content also have the highest LoS velocities. On the other hand,  gas particles with relatively lower \al{} content (i.e., not bright enough to be observed)  generally display the lowest LoS velocities. However, this result is not conclusive because the overall rotational velocity of our simulated galaxy is higher than that observed for cold CO gas in the Milky Way Galaxy. Furthermore, we found no significant correlation between gas temperature, rotational velocity, and \al{} content at any given radius. We also found the presence of transient \al{}-rich spots at low LoS velocities and we show that one such spot had been captured by the \spi{} data. Based on our model, we present a prediction for the detection of 1.8~MeV~$\gamma$-rays by the future COSI mission. We find that according to our model, the new instrument will be able to observe similar \al{}-emission patterns to those seen by  \spi{}.}  

{}


   \maketitle
%
%-------------------------------------------------------------------
\section{Introduction}
The short-lived radioactive isotope  \al, with a  a half-life of 0.72~Myr, is primarily produced and ejected by massive stars and their core-collapse supernovae (CCSNe). The detection of the 1.8 MeV~$\gamma$-ray produced by its decay makes it a tracer of active nucleosynthesis and star formation in the Galaxy \citep{Prantzos96,Diehl23} and a means to study dynamics and feedback processes in the interstellar medium \citep[ISM,][]{Diehl06,Diehl10,Wang09,Kretschmer13,Bouchet15,Siegert17}. One main observational feature of such detection, however, remains a topic of speculation and debate: when plotted according to their location and line-of-sight (LoS) velocity relative to the Solar System, the $\gamma$-ray flux tracing the \al{} content, is higher for higher LoS velocities. Specifically, the LoS velocity of the \al{}-rich regions reaches values higher than those observed with CO. To resolve this tension, several models and scenarios have been discussed, which can roughly be divided into two groups. 

Previous works \citep{Krause18,Diehl10,Rodgers-Lee19} have suggested that the component with both a high LoS velocity and high \al{} content is caused by young star associations that migrate towards the front of spiral arms. When CCSNe occur in these associations and eject \al{} isotropically, the component of the produced \al{} traveling away from the spiral arm would have a higher velocity for an observer than the component traveling towards the inside of the spiral arm. This is due to the different gas density, as it would slow down the ejected \al{} \citep[see also, e.g.,][]{Diehl22,Diehl23}. Some of these models rely on the presence of long-lived, stationary, density wave-type spiral arms, leading to an asymmetry between the lead and tail side of the spiral arm. However, numerical, long-duration galaxy simulations have shown that spiral arms are rather short-lived (on the order of $100$~My) and non-stationary \citep{Wada11,Grand12,Baba13}.

The other approach suggests that the component with high LoS velocity and high \al{} content is due to foreground material. To achieve the high LoS velocities of \al{} seen in the \spi{} observations, it is required that we set the Solar System  in a superbubble with a fresh \al{} supply \citep[e.g.,][]{Fujimoto20}. This scenario initially appeared to be in agreement with the detection of fresh interstellar $^{60}$Fe and $^{244}$Pu in deep-sea sediments and Earth crust samples, however, it was  later shown that these isotopes might have originated from farther away and could have been brought to the Solar System by diffusion \citep{Hotokezaka15}. On this basis,  a local enriched bubble would not necessarily be required to explain these deep-sea detections \citep{Wehmeyer23}.

To revisit this problem, we used the cosmological, chemodynamical zoom-in simulation of a Milky-Way type galaxy described in \cite{Haynes19} and \cite{vincenzo20}, which includes star formation, feedback, and detailed chemical enrichment. It reproduces a number of observations in the Milky Way \citep[see][for a review]{kob23book}.

In this paper we include, for the first time, the production and decay of \al{}, fully self-consistently. We  also give a comparison of the locations and velocities of \al{}-rich gas particles in the simulated galaxy with the measurements of \spi{}. 

Our model represents a first-principle realisation of the \al{} content in the simulated galaxy, as it does not rely on assumptions about density wave-type spiral arms, fixed axisymmetric stellar and dark matter distributions \citep[as, e.g., done in][]{Fujimoto18}, stochastic star formation \citep[as, e.g., done in][]{Fujimoto20}, or snapshots of longer-lasting galaxy simulation models. Instead, we used a cosmological zoom-in model coupled with a state-of-the-art galactic chemical evolution (GCE) model to investigate the chemical and dynamical evolution of a simulated galaxy, whose star formation history is determined from cosmological accretion of gas, a series of minor mergers of satellite galaxies, and feedback from previous generation of stars.
In Section~\ref{Section:Methods}, we describe our method and the galaxy evolution model. In Section~\ref{Section:Results}, we present the model predictions, along with a comparison to the observational data and predictions for the detections of the future COSI mission. 

\section{Methods} 
\subsection{Cosmological zoom-in model} \label{Section:Methods}
We used a state-of-the-art GCE model coupled with the hydrodynamical code {\tt Gadget-3} \citep{spr05}. For all the stable elements, the code includes the most up-to-date nucleosynthesis yields from asymptotic giant branch (AGB) stars, which evolve from low-mass stars, core-collapse supernovae (CCSNe, including hypernovae, \citealt{Kobayashi20}), and Type Ia supernovae \citep{kobayashi20ia}.
The code was originally developed by \cite{Kobayashi04,kob07} and applied to simulate a Milky-Way-type galaxy, as described by \cite{Kobayashi11,Haynes19,vincenzo20}.
The details, including all the relevant mathematical formulae, can be found in \cite{kob23book}. In brief, the most relevant features of the model are:
\begin{itemize}
\item The gravitational forces are calculated with tree-particle-mesh scheme and hydrodynamics is followed according to the smoothed-particle-hydrodynamics (SPH) method \citep{spr05}.
\item The metal-dependent cooling functions are generated with the MAPPINGS III software \citep{Sutherland93}, assuming the observed [O/Fe]--[Fe/H] relation in the solar neighborhood.
\item The star formation criteria are (i) convergent, (ii) cooling, and (iii) Jeans unstable. For the star formation rate, we used a Schmidt law, namely, the star formation timescale is proportional to the dynamical timescale, $t_{\rm sf}  \equiv  t_{\rm dyn}/c$, where $c$ is a constant equal to $0.02$.
\item Once the star formation criterion is fulfilled in a gas particle, some of its mass converts into a star particle with a mass, $m$, between $10^5 \,\mathrm{M}_\odot \ltsim \, m \ltsim 10^6\,\mathrm{M}_\odot$. For the masses of individual stars in the star particles, we used the IMF from \cite{kro08}  (with the massive-end slope of $x=1.3$) in the mass range of $0.01 \,\mathrm{M}_\odot \leq m \leq 120 \,\mathrm{M}_\odot$. 
\item The energy and heavy element yields from dying star particles are distributed to a fixed number, $N_{\rm FB}=64,$ of neighbouring gas particles, weighted by the SPH kernel. Further metal diffusion is not included as no impact of it was found in \cite{Haynes19}.
The energy from stellar winds and supernovae (each between 1--30 $\times 10^{51}$ erg) is thermally distributed.
\end{itemize}

The cosmological initial conditions (at redshift $z=127$) are the same as in \cite{Haynes19,vincenzo20}. The initial mass of each gas particle is $3.3 \times 10^6 \,\mathrm{M}_\odot$ and the gravitational softening length is $1.4$ kpc.
The detailed star formation history and the spatial distributions of gas, stars, and metals can be found in \cite{Haynes19,vincenzo20}; there is the bulge, the thin and thick disks, and the halo, as in our Milky Way. The bulge is formed by an assembly of gas-rich sub-galaxies at $z\gtsim2$ and the majority of stars there are old, metal-rich, and $\alpha$-enhanced. The disk is formed inside-out and star formation takes place continuously, self-regulated by cosmological inflow and supernova-driven outflow, generating the decreasing trend of [$\alpha$/Fe] ratio toward lower metallicities, consistently with the observations.
\newline
The main difference between the \cite{Kobayashi11,Haynes19,vincenzo20} model and this work is, that we introduce radioactive nuclei as described in the following Section. Note that the introduction of these nuclei does not alter the evolution of the simulated galaxy (e.g., age-metallicity relation, star formation rate), except for the presence of these nuclei.

\subsection{Radioactive nuclei}

The stellar sources of \al{} have been extensively discussed in the literature \citep[see, e.g.,][for reviews]{Prantzos96,Diehl21,Laird23}. Here, we chose CCSNe from massive stars as the exclusive \al{} source. The literature also considers other sources of \al{}: (super-)AGB stars and novae. For AGB stars, \cite{Lugaro08} found that the initial mass function (IMF)-averaged contribution divided by the contribution of CCSNe is only on the order of $0.2 \%$. Super-AGB stars, which have significantly higher \al{} yields than AGB stars, still only contribute to about $10 \%$ of the Galactic inventory. For instance, \cite{Siess08}  also found that their impact on GCE would be insignificant. For novae, despite the fact that some studies \citep{Vasini22,Vasini25} have claimed a relative contribution to the Galactic inventory of up to $75 \%$, \cite{Laird23} pointed out that their contribution is subject to huge uncertainties regarding their progenitors (ONe white dwarfs versus ONeMg white dwarfs) and the metallicity range in which they would contribute. They find a find a Galactic contribution of only about $12 \%$ \citep[e.g.,][]{Laird23}. Furthermore, the observed \al{} distribution is consistent with massive stars being the dominant source of the Galactic \al{} \citep{Prantzos96,Diehl06}. Still, the unclear and recently newly discussed nova contribution to the Galactic \al{} inventory should be studied using a cosmological model in the future.
\newline
In our simulation, we used CCSNe in the mass range $13 \,\mathrm{M}_\odot \leq \, m \leq 40\,\mathrm{M}_\odot$, with the yields from \cite{Nomoto13}, and AGB stars, with the yields from Table~S1 in \cite{Lugaro14} as \al{} sources. Averaging the \cite{Nomoto13} \al{} yields over the Salpeter IMF (with a constant slope $-1.35$), the average \al{} yield is $\approx$$2.90 \times 10^{-5}\mathrm{M}_\odot$ ($\approx$$1.03 \times 10^{-6}\mathrm{M}_\odot$ for \al{} from AGB stars).
For comparison, \cite{Fujimoto20} used the \cite{Sukhbold16} yields, but multiplied by a factor of two (to ensure that the steady-state mass ratio of $^{60}$Fe/\al{} is consistent with observations), which corresponds to a yield of 2.72--8.44~$\times 10^{-5}\mathrm{M}_\odot$ of \al{}, depending on the supernova engine.
\cite{Rodgers-Lee19} and \cite{krause21} injected a passive scalar tracer fluid evolving according to the \cite{Voss09} population synthesis model every time a superbubble forms and used the yields from \cite{Limongi06a} or \cite{Woosley95}, which give the IMF-averaged \al{} yield of $\approx$$6.17 \times 10^{-5}\mathrm{M}_\odot$, or $\approx$$5.84 \times 10^{-5}\mathrm{M}_\odot$, respectively.
\cite{Pleintinger19} used a population synthesis code (Stochastically Lighting Up Galaxies, SLUG; e.g., \citealt{Dasilva12,Krumholz15}) with the \cite{Sukhbold16} yields (see above numbers, but noting that in that study, they did not multiply by a factor of two).
For \cite{Vasini25}, in their Model-1, as \al{} sources, they used low mass ($m \leq 6\,\mathrm{M}_\odot$) AGB stars (yields from \citealt{Karakas10}), SNeIa (\al{} yields from \citealt{Nomoto84}), and CCSNe (yields from \citealt{Woosley95}). 
Despite the rather high averaged \al{} CCSN yield, this model had difficulties reaching the observed galactic \al{} content. Hence, they introduced their Model-2, with novae (\al{} yield $\approx$$6.17 \times 10^{-5}\mathrm{M}_\odot$ of \al{} per event, according to the simulations of \citealt{Jose07}, with the occurrence considerations in \citealt{Romano03,Vasini22}) as an additional \al{} production site.

In our simulation, \al{} produced by CCSNe in star particles is distributed in the same way as stable nuclei to neighboring gas particles. To determine the abundance of \al{} (i.e., $Y_{26}$) in a particle in the current time step ($\text{t}+\Delta \text{t}$), we use the exponential decay law $Y_{26}(\text{t}+\Delta \text{t})=Y_{26}(\text{t}) \exp (-\Delta \text{t}/\tau_{26})$, with $\tau_{26}$(=1.035 Myr) being the mean life of \al{}, and $Y_{26}(\text{t})$ the abundance of \al{} in the gas particle in the previous time step. The heating from radioactive decay of \al{} is negligible for galaxy evolution.

\section{Results}
\label{Section:Results}

We present our results in terms of \al{} content, locations, and velocities. The different locations and velocities used below are summarized in Table~\ref{Table1}, together with their definition and where we  used them to illustrate the results.

\subsection{\al{}-distribution in the galaxy}
\begin{figure}
    \centering
    \includegraphics[width=\columnwidth]{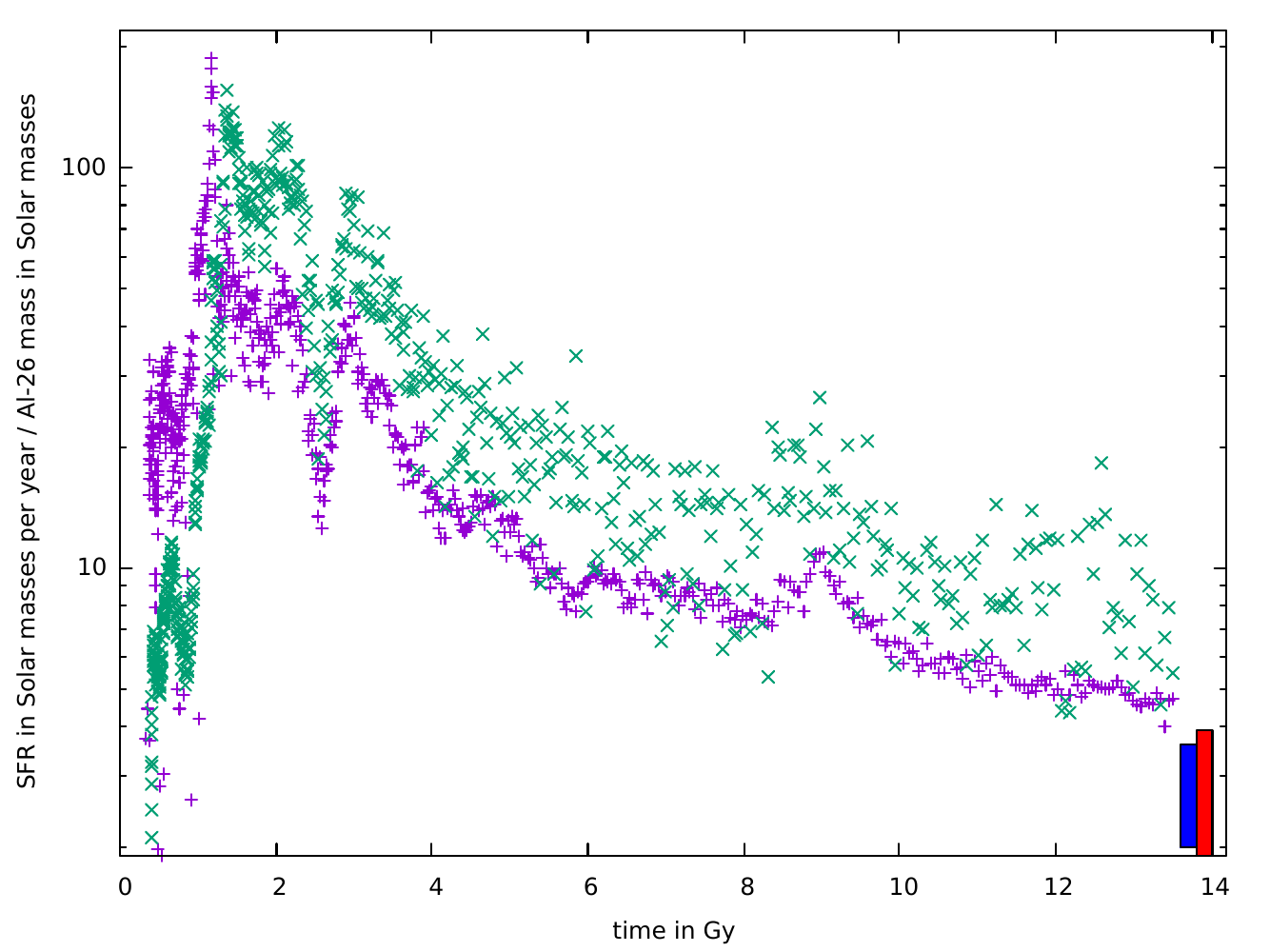}
    \caption{Star formation rate in solar masses per year (magenta crosses), and total \al{} mass in solar masses (green crosses) in the simulated galaxy as a function of galactic evolution time. Today's values for the total \al{} content in the Galaxy deduced from the \spi{} observations ($2.8 \pm 0.8$ M$_\odot$, \citealt{Diehl06}) are indicated by a (time-offset) blue box, and estimates for the current star formation rate in the Galaxy ($0.35-3.9 $ M$_\odot$ per year, Table~1 in \citealt{Elia22}) are indicated by a (time-offset) red box. Note: the red box continues below the figure.}
    \label{fig:Masses}
\end{figure}
Figure~\ref{fig:Masses} shows the time evolution of the star formation rate and the total \al{} mass contained in the simulated galaxy. At $t\gtrapprox$ 6 Gyr, the \al{} content in the galaxy has settled into a steady state. In the equilibrium state, the star formation rate fluctuates between roughly one solar mass per year up to several hundreds of solar masses per year, while the total \al{} content in the galaxy fluctuates between roughly 3 and 30 solar masses of \al{}. The lower limit of these fluctuations is consistent with the estimate of 1.7M$_\odot \leq$ m$_{26}$ $\leq$ 3.5 M$_\odot$ in \citealt{Diehl21} and  slightly more than the best estimate of 2M$_\odot$ of \al{} in \citealt{Pleintinger20} and the calculations of \cite{Fujimoto20} for the galactic disk ($\approx$1.1M$_\odot$ of \al{}).

\begin{figure}
    \centering
    \includegraphics[width=\columnwidth]{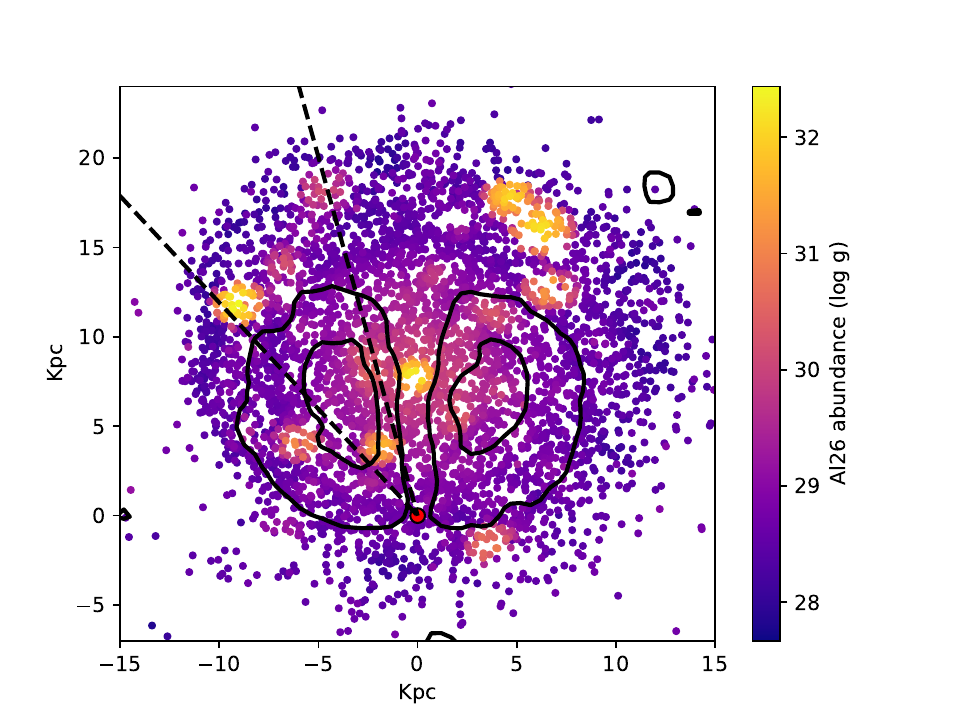}
    \caption{The \al{} distribution in our simulated galaxy. The dots represent disk gas particles at present-day, color-coded to represent the mass of \al{} (in log(gram) unit). The red circle at coordinates [0:0:0] denotes the location of the observer (Solar System) at $R_\odot=8.5$~kpc from the galactic center. The black dashed lines indicate a viewing angle of $-40 \leq \phi \leq -14$, corresponding to the viewing angle highlighted by the black dashed vertical lines in Figure~\ref{fig:Al26v}. The contour plots represent the LoS velocity to the observer. The outer (inner) black contour represents the $\pm100$km/s ($\pm200$km/s) threshold. The single particles in the external regions with velocities higher than $\pm100$km/s originate from an earlier dwarf galaxy disruption event.}
    \label{fig:xy}
\end{figure}

\begin{table*}[]
    \centering
    \begin{tabular}{llll}
    \hline \hline
    Name &Symbol&Components/Calculation&Usage\\
    \hline
    
    Location & $\overrightarrow{x}$& [$x$:$y$:$z$] &  \\

Velocity& $\overrightarrow{v}$& [$v_x$:$v_y$:$v_z$] &  \\

    Location of the Solar System (observer)&$\overrightarrow{x_\odot}$&[0:0:0]& Section 3\\
%    &&&\\
        Velocity of the Solar System (observer)&$\overrightarrow{v_\odot}$&[+251 Km/s:0:0]& Section 3\\
%    &&&\\
    Location of the galactic center&$\overrightarrow{x_\text{GC}}$&[0:+8.5 Kpc:0]&Section 3\\
%    &&&\\

         &&&\\
            Line-of-sight (LOS) velocity &$v_\text{LOS}$         & $\frac{(v_x - 251 \text{Km/s}) \, x \,+\, v_y \,  y \,+\, v_z \, z}{\sqrt{x^2+y^2+z^2}}$  & Figs. 2, 3, 4 \\
          &&&\\
%         &&&\\
            Rotational velocity component&$v_\text{rot}$         & $\frac{-v_x \, (y - 8.5 \text{Kpc})\, + \, v_y \, x}{\sqrt{x^2+(y - 8.5 \text{Kpc})^2+z^2}  \sqrt{(v_x^2+v_y^2+v_z^2}}$  &  \\
     &&& $v_\text{rot}$/$\sigma >0.9$ determines if \\
    Velocity dispersion& $\sigma$     &$\frac{m \, [-v_x \, (y - 8.5 \text{Kpc})\, + \, v_y \, x]}{\sqrt{x^2+(y - 8.5 \text{Kpc})^2+z^2}}$    &the particle is a disk particle\\
    \hline
    \end{tabular}
    \caption{Overview of the different locations and velocities used in the text and figures.}
    \label{Table1}
\end{table*}
Figure~\ref{fig:xy} shows the present-day distribution of the disk's gas particles in our simulation, color-coded according to their \al{} content. The figure also shows the contours of the LoS velocity of the particles, relative to the observer (Solar System) located at $\overrightarrow{x_\odot}$=[0:0:0]. This helps us to identify the location of the particles with the fastest LoS velocity.
In the galactic center, where the star formation activity is the highest, a highly (relatively homogeneously) \al{}-enriched region is found.
Outside of the galactic center, scattered \al{}-enriched regions are found. The cores of such scattered \al{}-bright spots host young star particles. Given the short life of massive stars, these particles feature a strong supernova (and thus \al{} production) activity. This leads to a high \al{} abundance in the surrounding  ISM of these star particles. New stars that form in these regions of the ISM also have a high \al{} abundance. As time progresses, the supernova activity in the dying star particles decreases, the radioactive decay of \al{} dominates over its production and the amount of  \al{} decreases  in both the star particles and in the surroundings.

Therefore, the \al{}-bright spots outside the galactic center are transient and disappear after a few Myr.
We note that this clumpy structure is inherent to choosing massive stars as the only source of galactic \al{}. If asymptotic giant branch stars (or novae) were instead the dominant source, we would instead expect a much smoother distribution because these stars (i) have a much wider distribution of possible lifetime expectancies, which would allow them to produce and eject \al{} over much longer time spans and counteract radioactive decay over longer time scales; and (ii) tend to be more homogeneously distributed throughout the galaxy. 
This more even distribution of \al{} sources would then lead to a much more uniform \al{} distribution in the galaxy.

\subsection{\al{}-longitude-LoS velocity diagram}
\begin{figure}
    \centering
    \includegraphics[width=\columnwidth]{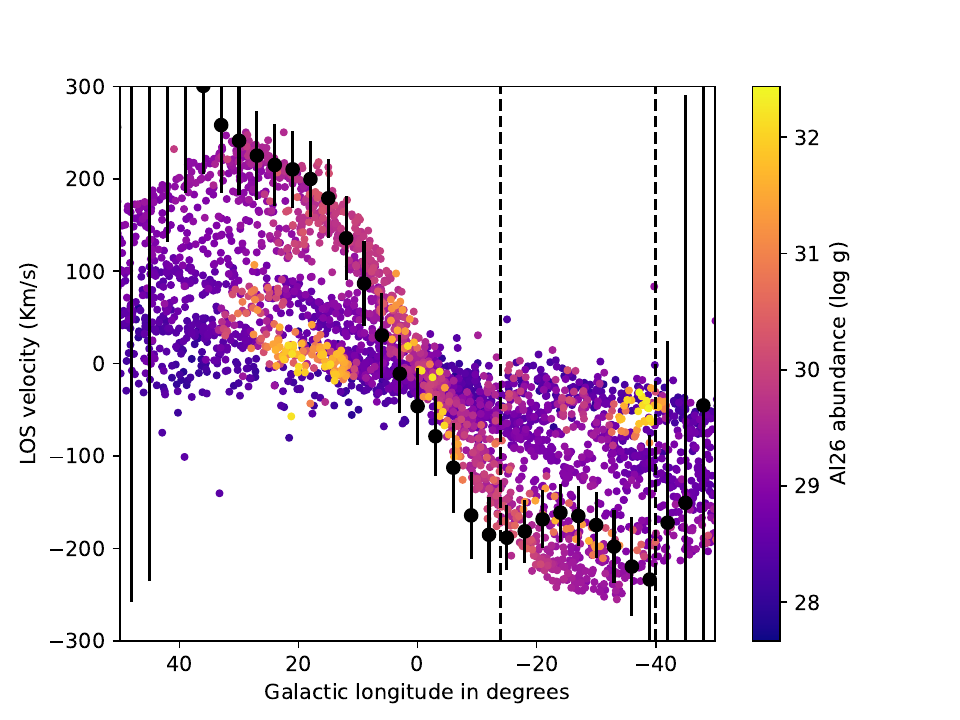}
    \caption{\al{}-longitude-LoS velocity diagram of the same disk gas particles of Figure~\ref{fig:xy}, as seen from the location of the Solar System (i.e., the red circle in Figure~\ref{fig:xy}), color-coded according to their \al{} content (in log(gram) unit). 
    Black dots with error bars show the \al{} detection data by the \spi{} instrument \citep{Kretschmer13} and Karsten Kretschmer, personal communication. 
    The dashed vertical lines highlight the region between the viewing angle $-40 \leq \phi \leq -14$, corresponding to the dashed lines in Figure~\ref{fig:xy}, showing the deviation discussed in the text.}
    \label{fig:Al26v}
\end{figure}

Figure~\ref{fig:Al26v} shows the \al{} content and LoS velocity of the simulated disk gas particles.
We define the LoS velocity as the velocity of a simulated particle relative to the Solar System following the rotation of the simulated galaxy (counter-clockwise in the framework of Figure~\ref{fig:xy}).

The simulated particles are compared to the data from the SPI detector on the INTEGRAL spacecraft, which was designed to detect the 1809~keV line from the decay of \al{} with an energy resolution of 3~keV. We plotted the \al{} decay $\gamma$-ray detection data by the \spi{} spacecraft \citep{Kretschmer13} as black error bars in Figure~\ref{fig:Al26v}. Contrary to the approach of \cite{Kretschmer13}, we plotted the full distribution of detection angles (Karsten Kretschmer, personal comm.) to highlight the small deviation from the main detected trend between $-40 \leq \phi \leq -14$, which we discuss in more detail below. We note that this choice results in two neighboring detections not being statistically independent, because each detection overlaps the neighboring one by 75\%.

It can be seen that, generally, the particles in the simulation that feature the highest \al{} mass (brightest colors) overlap with the \cite{Kretschmer13} observations in that they also appear to have the highest LoS velocities (except for the few bright spots discussed below). Instead, the particles that feature lower \al{} mass do not overlap with the observations, as expected, since the low \al{} abundances are more difficult to detect and generally show lower LoS velocities. 
The \al{}-rich particles in the outer regions of the simulated galaxy do not strongly affect the diagram of Figure~\ref{fig:Al26v}, as they are located in areas that do not feature a high relative velocity with respect to the observer (see Figure~\ref{fig:xy}). The \al{}-rich areas within the high-velocity contours are instead located close by. 

We note that our cosmological initial conditions result in a galaxy that is generally rotating faster than the Milky Way \citep{Scannapieco12}.
Although molecular clouds are not well resolved with the resolution of our simulation, the cold gas is rotating faster than the CO observation \citep[e.g.,][]{Sofue09}.
We do not see a significant correlation between gas temperature, rotational velocity, and \al{} content, 
at any given radius.
Our simulation self-consistently produces the formation of the Galactic bulge. This greatly affects the shape of the rotational velocity, resulting in a fast increase in the rotational velocity from the center to a few kpc outwards, before showing a flat rotation curve. The shape of the observed \al{}-longitude-LoS velocity diagram is strongly determined by the velocity and \al{} content of the material near the bulge (Figure~\ref{fig:xy}), which is self-consistently modelled in our simulation.

Figure~\ref{fig:Al26v} also shows a few \al{}-rich spots at low LoS velocities.
The origin of these \al{}-rich spots can be identified by correspondence to the transient \al{}-rich spots seen in Figure~\ref{fig:xy}. 
One such \al{}-rich spot reproduces the observed distinct deviation in the \spi{} detections at a viewing angle of $-40 \leq \phi \leq -14$. In Figure~\ref{fig:xy}, the corresponding \al{}-rich spot is the one closest to the observer in this viewing angle, at the edge of the $-200$~km/s contour.
This \al{}-rich spot, which matches the observed deviation is located at the edge of the $-200$km/s contour in Figure~\ref{fig:xy}. Inside the contour, there are particles that rotate faster, but have lower \al{} abundance than the \al{}-rich spot. This contrast causes the deviation observed by \spi{}. The other two predicted \al{}-rich spots were not detected by SPI; however, this was expected, as they are transient. On the other hand,  Figure~\ref{fig:Masses} shows that the current galactic period features below-average \al{}-content compared to the previous $\approx$4~Gy of evolution.

\begin{figure}
    \centering
    \includegraphics[width=\columnwidth]{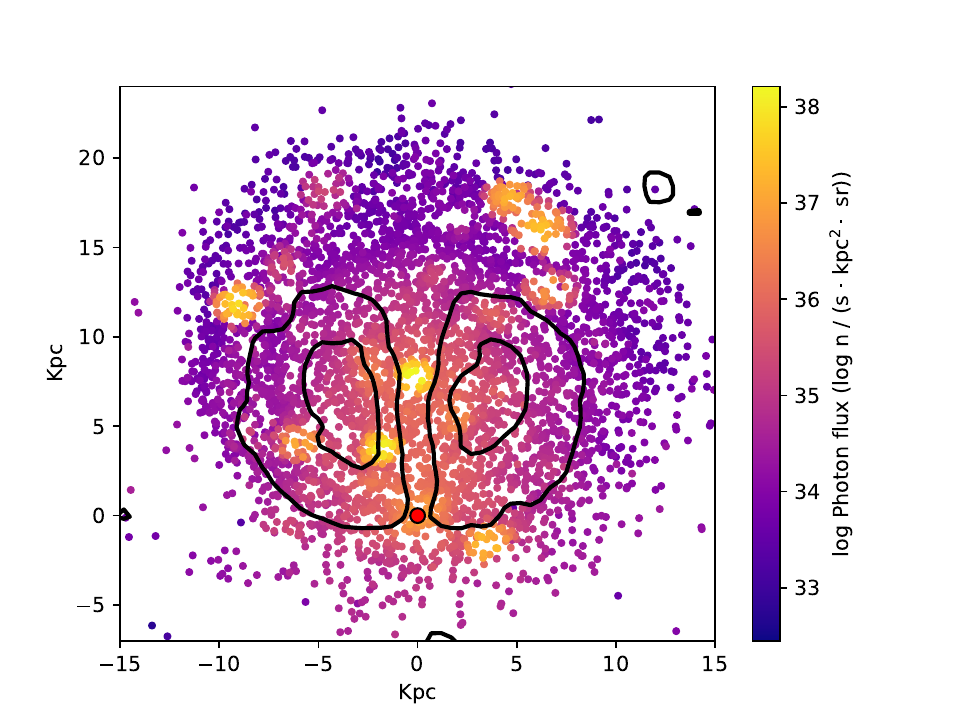}
    \caption{Same as Figure~\ref{fig:xy}, but color-coded as function of 
    the number of emitted 1.8~MeV $\gamma$-rays per second by each gas particle, as seen by the observer. The outer (inner) black contour representing the $100$km/s ($200$km/s) threshold (as in Figure~\ref{fig:xy}) are retained here as a visual guide.} 

    \label{fig:Al26dilutionvelocity}
\end{figure}

\subsection{Future missions}
\label{Sec:Results:Future missions}

The upcoming COSI mission will perform a full sky survey, collecting many more photos than \spi{} and thus providing a full galactic map with smaller uncertainties. This will be possible thanks to its field of view of 25\% of the sky (relative to 1\% of \spi{}), along with a similar spectral resolution and angular resolution of $1.8^{\circ}$(compared to $2.7^{\circ}$) to \spi{}. 

\begin{figure}
    \centering
    \includegraphics[width=\columnwidth]{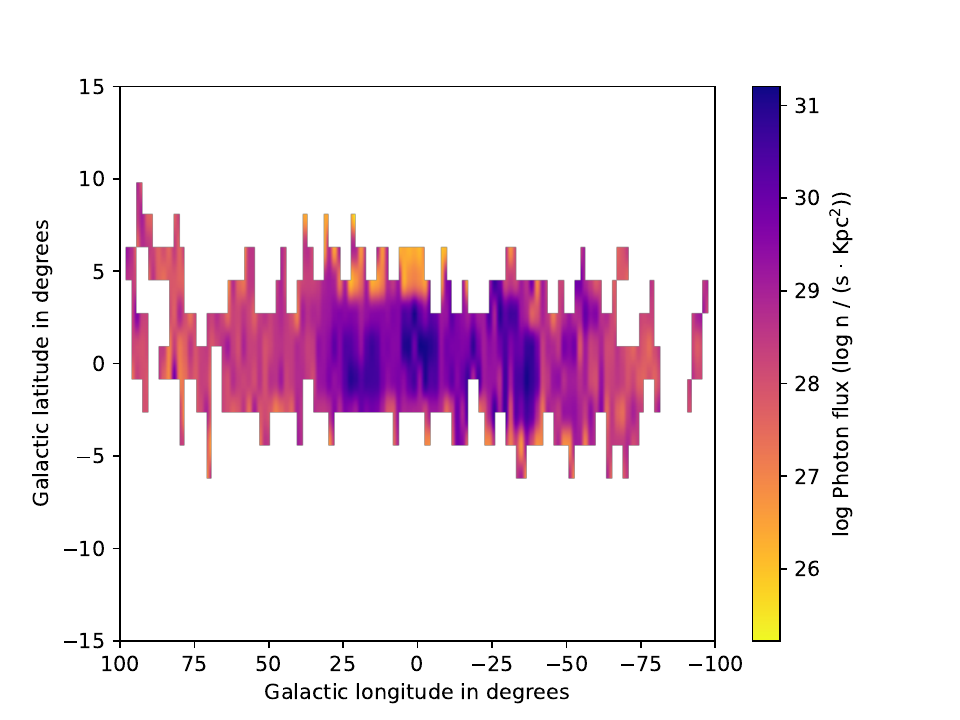}
    \caption{Prediction for possible detection of \al{}-decay $\gamma$-ray as seen by the future COSI $\gamma$-ray detection mission.
    The x- and y- axis represent the viewing angles of the future instrument. Note: the color-coding here is opposite to that in Figs.~1~and~2 (to achieve a similar color-coding as in \citealt{Kretschmer13}), so that higher intensity is represented by darker colors and no detection is represented by the white background. 
The pixels are Kaiser-interpolated to account for a possible neighbour dependence, as in \spi{}.}
    \label{fig:Al26Instr}
    \end{figure}

We analysed our simulations using the angular resolution of $1.8^{\circ}$ of the future $\gamma$-ray detection mission COSI \citep{Tomsick23}. To achieve a view of our simulated galaxy as this future instrument will see it, we used the following method:
\begin{enumerate}
\item We take the spatial distribution of the gas particles from the simulation (Figure~\ref{fig:xy}) and multiply the \al{} mass by the Avogadro number, then divide by its atomic mass 26 to obtain the number of \al{} atoms in the gas particle.
\item We multiply the number of \al{} atoms by the \al{} decay rate of $9.68 \times 10^{-7}$~s$^{-1}$ to find the number of 1.8 MeV $\gamma$-rays emitted per second by each gas particle.
\item We divide this number by the square root of the distance to the observer to account for the decrease in $\gamma$-ray counts with increased distance from the source. The result is shown in Figure~\ref{fig:Al26dilutionvelocity}. In the figure, we can see which parts of the simulated galaxy contribute the most 1.8~MeV $\gamma$-ray intensity towards an observer located at the location of the Solar System.

\item We sum up all the contributions located in each viewing cone from the observer, binned according to the resolution of the COSI instrument of $(4.4^{\circ})^2$, as shown in Figure~\ref{fig:Al26Instr}. 
\end{enumerate}

We predict that the instrument will receive the highest intensity of \al{}-decay $\gamma$-rays from the direction of the galactic plane. The majority of $\gamma$-rays, however, do not appear to originate from throughout the galactic plane uniformly, but their predicted distribution appears to be clumped, similarly to  detections carried out by \spi{} \citep[see Figure~1 of][]{Kretschmer13}. This clumpiness originates from the \al{}-rich spots present throughout the galaxy, which is already considered as a key piece of  evidence that low-mass AGB stars and novae are not the dominant source of \al{} in the Galaxy, as they would produce a more uniform distribution of $\gamma$-ray sources and, therefore, of the intensity as well.

\section{Discussion and conclusions}

We have run, for the first time, a cosmological zoom-in chemodynamical simulation of a Milky-Way-type galaxy including radioactive nuclei to predict the distribution of \al{} from massive stars in a simulated galaxy. Our main conclusions are as follows:
\begin{itemize}

    \item The \al{} distribution in the simulated galaxy appears clumped, rather than heterogeneous.
This is a consequence of the fact that our model only includes CCSNe from massive stars as the source of \al{}, in agreement with previous observations by the \spi{} spacecraft \citep[][]{Kretschmer13} and theoretical considerations \citep{Prantzos96,Wang09,Bouchet15,Diehl23}. Massive stars are associated with young star clusters, which are distributed in a clumpy way throughout the galaxy.  

    \item The \al{}-rich regions generally appear at the highest LoS velocities to an observer located in the position of the Solar System. 
    This in agreement with \al{} decay observations by the \spi{} spacecraft \citep[][]{Kretschmer13} and confirms the results from the model of \citet{Fujimoto20}, for an observer outside of \al{} bubbles.

\item    Transient \al{}-rich spots of lower LoS velocities are present in the simulation and one of them appears to be coincident with a deviations observed in the \spi{} detection.

\item We also illustrated our results as they would be observed by the future observation mission COSI.
From our modeled galaxy, COSI would detect a similar, clumped \al{}-emission pattern as \spi{}. This is due to \al{} predominantly produced in massive stars, which are spatially distributed in a less uniform way as intermediate mass stars.

\end{itemize}

Our model can trace the radioactivity in the galaxy following the structure formation, star formation, and chemical enrichment histories in a cosmological context. We showed that the observation can be well reproduced if the hot, \al{}-rich gas is rotating as fast as in our simulated galaxy with a clumpy distribution. As noted earlier in this work, however, our simulated galaxy is not exactly the same as our Milky Way Galaxy; specifically, it has a faster rotational velocity. Therefore, more simulations are required to investigate the behaviour of a galaxy with more similar properties to the Milky Way.

Future works will include more isotopes of interest, for example, $^{60}$Fe and $^{244}$Pu, and should also integrate more detailed propagation of radioactive isotopes, either in the form of gas or dust \citep[e.g.,][]{Hotokezaka18}. Furthermore, the unclear role of novae \citep{Vasini22,Laird23,Vasini25} and their implications for the \spi{} detections should be addressed in a future cosmological model as well.

\begin{acknowledgements}
The authors thank Karsten Kretschmer for providing the detection data relevant for this work, Thomas Siegert for providing insights into the COSI instruments, and Roland Diehl, and Martin G. H. Krause for many useful discussions.
BW acknowledges support from the National Center for Science (NCN, Poland) under Grant no. 2022/47/D/ST9/03092, as well as the National Science Foundation (NSF, USA) under Grant no. PHY-1430152 (JINA Center for the Evolution of the Elements) and Grant no. OISE-1927130 (IReNA). 
CK acknowledges funding from the UK Science and Technology Facility Council (STFC) through grant ST/R000905/1, \& ST/V000632/1.
The work was also funded by a Leverhulme Trust Research Project Grant on ``Birth of Elements''. 
This work was supported by the ERC Consolidator Grant (Hungary) funding scheme (Project RADIOSTAR, G.A. no. 724560) and by the Lend\"ulet Program LP2023-10 of the Hungarian Academy of Sciences. We also thank the COST actions ``ChETEC'' (G. A. no. 16117) and ``ChETEC-INFRA'' (G. A. no. 101008324). M.L. was also supported by the NKFIH excellence grant TKP2021-NKTA-64. 
The work of AYL was supported by the US Department of Energy through the Los Alamos National Laboratory. Los Alamos National Laboratory is operated by Triad National Security, LLC, for the National Nuclear Security Administration of U.S.\ Department of Energy (Contract No.\ 89233218CNA000001).
Some computations outlined in this paper were performed at the Wroclaw Centre for Scientific Computing and Networking (WCSS).
We thank an anonymous reviewer for their valuable suggestions improving the manuscript.
\end{acknowledgements}

\bibliographystyle{aa}
\bibliography{main}

\end{document}